\documentclass[10pt,aps,prc,twocolumn,superscriptaddress,floatfix]{revtex4-1}
\usepackage{graphicx}
\usepackage{amsmath}
\usepackage{braket}
\usepackage{url}

\usepackage{multirow}
\usepackage{amssymb} 
\usepackage{booktabs}
\usepackage{rotating}
\usepackage{bm}
\usepackage{bbm}
\usepackage{longtable}
\usepackage{subfigure}
\usepackage[colorlinks,
linkcolor=blue,
anchorcolor=red,
urlcolor=red,
citecolor=red]{hyperref}

\usepackage[utf8]{inputenc}
\usepackage[T1]{fontenc}

\begin{document}
%\begin{CJK}{UTF8}{gbsn} 

\title{Gamow shell model study of the $^{17}$Ne$(p,p)$ reaction and of isospin symmetry breaking in $^{18}$Na}

\author{N. Chen}
\affiliation{State Key Laboratory of Heavy Ion Science and Technology, Institute of Modern Physics, Chinese Academy of Sciences, Lanzhou 730000, China}
\affiliation{School of Nuclear Science and Technology, University of Chinese Academy of Sciences, Beijing 100049, China}

\author{J. G. Li}\email[]{jianguo\_li@impcas.ac.cn}
\affiliation{State Key Laboratory of Heavy Ion Science and Technology, Institute of Modern Physics, Chinese Academy of Sciences, Lanzhou 730000, China}
\affiliation{School of Nuclear Science and Technology, University of Chinese Academy of Sciences, Beijing 100049, China}
\affiliation{Southern Center for Nuclear-Science Theory (SCNT), Institute of Modern Physics, Chinese Academy of Sciences, Huizhou 516000, China}

\author{K. H. Li}
\affiliation{College of Physics, Henan Normal University, Xinxiang 453007, China}
\affiliation{State Key Laboratory of Heavy Ion Science and Technology, Institute of Modern Physics, Chinese Academy of Sciences, Lanzhou 730000, China}

\author{N. Michel}\email[]{nicolas.michel@impcas.ac.cn}
\affiliation{State Key Laboratory of Heavy Ion Science and Technology, Institute of Modern Physics, Chinese Academy of Sciences, Lanzhou 730000, China}
\affiliation{School of Nuclear Science and Technology, University of Chinese Academy of Sciences, Beijing 100049, China}

\author{P. Y. Wang}
\affiliation{State Key Laboratory of Heavy Ion Science and Technology, Institute of Modern Physics, Chinese Academy of Sciences, Lanzhou 730000, China}
\affiliation{School of Nuclear Science and Technology, University of Chinese Academy of Sciences, Beijing 100049, China}

\author{W. Zuo}\email[]{zuowei@impcas.ac.cn}
\affiliation{State Key Laboratory of Heavy Ion Science and Technology, Institute of Modern Physics, Chinese Academy of Sciences, Lanzhou 730000, China}
\affiliation{School of Nuclear Science and Technology, University of Chinese Academy of Sciences, Beijing 100049, China}
\affiliation{Southern Center for Nuclear-Science Theory (SCNT), Institute of Modern Physics, Chinese Academy of Sciences, Huizhou 516000, China}

%\affiliation{Institute of Modern Physics, Chinese Academy of Sciences, Lanzhou 730000, China}
%\affiliation{School of Nuclear Science and Technology, University of Chinese Academy of Sciences, Beijing 100049, China}
%\email[]{nicolas.michel@impcas.ac.cn}
\date{\today}

\begin{abstract}
The unbound nucleus $^{18}$Na, acting as an intermediate nucleus in the sequential decay of $^{19}$Mg, is situated beyond the proton drip line. We employ the coupled-channel Gamow shell model (GSM-CC) to investigate the properties of $^{18}$Na, as well as the $^{17}\text{Ne}(p,p)$ cross section. 
GSM-CC treats the nucleus as an open quantum system and provides a unified framework for studying both nuclear structure and reaction cross sections.
Our calculations reproduce the energies and partial decay widths of low-lying states in $^{18}$Na, as well as the $^{17}\text{Ne}(p,p)$ cross section.
Additionally, the mirror nucleus $^{18}$N is also described. The isospin symmetry breaking induced by the Coulomb interaction and continuum coupling is clearly obtained in our description of $^{18}$Na and $^{18}$N properties, arising from the extended $s_{1/2}$ partial wave. 
The isospin symmetry breaking in the $^{18}$Na/$^{18}$N pair is compared to that occurring in the mirror pair $^{16}$F/$^{16}$N, whereby similarities and differences are analyzed.

\end{abstract}

%\keywords{}

\maketitle

\section{Introduction}
In nuclei near or beyond drip lines, the nuclear force can hardly bind all nucleons together cohesively, so that nuclear systems become weakly bound or even unbound~\cite{Baumann2007,PhysRevLett.108.142503,PhysRevC.85.034327,PhysRevC.102.044614}.
Nuclei with extreme $N/Z$ ratios exhibit many properties that stable nuclei do not possess, such as particle emission~\cite{PhysRevLett.99.192501,Li2024,RevModPhys.84.567}, halo~\cite{TANIHATA2013215,PhysRevC.101.031301,RIISAGER1992393}, clusterization~\cite{VONOERTZEN200643,RevModPhys.90.035004}, Borromean structure~\cite{MA2020135673,RevModPhys.76.215}, and large mirror-symmetry breaking~\cite{PhysRevLett.133.222501,LI2023138197}. 
These phenomena mainly occur due to the strong continuum coupling and inter-nucleon correlations in drip-line nuclei, where valence nucleons primarily occupy orbitals near particle-emission threshold~\cite{DOBACZEWSKI2007432,SORLIN2008602}. 

The proton drip line is accessible except for the heaviest proton-rich systems and has generated significant interest.
As the number of protons in a proton-rich nucleus increases, the ground state separation energy of the nucleus will decrease, eventually causing the nucleus to become unbound and decay via proton emission, for example the one-proton ($1p$) decay observed in $^{15}$F~\cite{PhysRevC.69.031302} and $^{21}$Al~\cite{PhysRevC.110.L031301}.
$2p$ decay has been uncovered in a few situations, where initial nuclei remain bound against single-proton emission but directly emit two protons, such as: $^{45}$Fe~\cite{PhysRevLett.99.192501}, $^{54}$Zn~\cite{PhysRevLett.94.232501}, and $^{19}$Mg~\cite{PhysRevLett.99.182501}. 
Other various proton-emission modes have also been observed, such as $3p$ decay of $^7$B~\cite{PhysRevC.84.014320} and $^{13}$F~\cite{PhysRevLett.126.132501}, $4p$ decay of $^{8}$C~\cite{PhysRevLett.32.1207,PhysRevC.82.041304} and $^{18}$Mg~\cite{PhysRevLett.127.262502,PhysRevC.103.044319,Zhou2024}, and $5p$ decays of $^9$N~\cite{PhysRevLett.131.172501}. These exotic proton emission modes offer excellent laboratories for the study of nuclear interactions, nuclear correlations, and continuum coupling in complex nuclear systems.

Near the proton drip line, $^{17}$Ne is a candidate nucleus for a two-proton halo structure. Adding two protons to $^{17}$Ne produces the two‑proton emitter $^{19}$Mg, which decays either sequentially via the unbound intermediate nucleus $^{18}\text{Na}$ or through direct $2p$ emission. In both cases, the structure of the intermediate nucleus \(^{18}\)Na is essential and is accessible experimentally. %it is worthy of in-depth theoretical study for the intermediate nucleus $^{18}$Na which is accessible experimentally.
%$^{17}$Ne, near the proton drip line, is a candidate for a two-proton halo structure, often described as a Borromean system of 15O+p+p, where all binary subsystems are unbound. the two-proton Borromean halo nucleus 17Ne (15O+p+p). 17Ne—a Borromean three‐body system (15O+p+p)
%As an intermediate nucleus in the sequential decay process starting from $^{19}$Mg, the unbound nucleus $^{18}$Na happens to be accessible experimentally, so that it is worthy of in-depth theoretical study. Moreover, the theoretical calculations of $^{19}$Mg heavily rely on the structure of $^{18}$Na and its mirror nuclei~\cite{VONOERTZEN200643}. 
%The $^{18}$Na was first studied by means of stripping reaction in Ref.~\cite{Zerguerras2004}, providing ambiguous data of the ground state. Subsequent experimental discoveries have identified the g.s. and low-lying states of $^{18}$Na~\cite{PhysRevC.85.044325,ASSIE2012198}.
The first measurement of the one-proton decay of \(^{18}\text{Na}\) was observed in Ref.~\cite{Zerguerras2004} through the invariant-mass spectrum. However, due to ambiguous data, the decay scheme of this nucleus could not be fully determined. Subsequently, a \(^{17}\text{Ne}(p,p)\) resonant elastic scattering experiment was conducted~\cite{ASSIE2012198}, whereby the spectroscopic properties of the low-lying states were obtained through an \(R\)-matrix analysis. Additionally, \(^{18}\text{Na}\) was investigated by measuring the trajectories of its decay products using microstrip detectors~\cite{PhysRevC.85.044325}, leading to the detection of its ground state and the determination of its one-proton decay energy at 1.23(15) MeV. Furthermore, the invariant-mass method was employed to investigate the \(d\)-wave resonances of \(^{18}\text{Na}\)~\cite{PhysRevC.107.054301}.
Theoretical shell model (SM) calculation and a weak-coupling procedure are also used to calculate mass excess of $^{18}$Na~\cite{PhysRevC.73.064310,PhysRevC.76.014313,PhysRevC.85.051302}.

%To further explore the structure and decay properties of $^{18}$Na, we perform a detailed investigation of the $^{17}$Ne(p, p) reaction.

%As an intermediate nucleus in the decay process starting from $^{19}$Mg, the unbound nucleus $^{18}$Na is worthy of in-depth theoretical study. 
The $^{17}\text{Ne}(p, p)$ reaction serves as a precision probe in as the differential cross-sections issued from proton scattering angles allowing to identify resonance states in $^{18}\text{Na}$. Combined analyses of resonance energies, decay widths, configurations and spectroscopic factors then clarify the nuclear structure of $^{18}\text{Na}$.
Moreover, the mirror pair formed by $^{18}$Na (unbound) and $^{18}$N (deeply bound) is of particular interest, as nearby analogous systems, such as $^{16}$F/$^{16}$N and $^{20}$Al/$^{20}$N, exhibit pronounced isospin symmetry breaking~\cite{PhysRevC.106.L011301,hkmy-yfdk,PhysRevC.96.054316}.

For nuclei near drip lines, continuum coupling effects and inter-nucleon correlations become significant, whose intertwined features are lacking in the traditional harmonic oscillator shell model. Conversely, the Gamow shell model (GSM)~\cite{PhysRevLett.89.042502,PhysRevLett.97.110603,PhysRevLett.127.262502,Michel2009,PhysRevC.84.051304,PhysRevC.96.054316,PhysRevC.67.054311,PhysRevC.103.034305,PhysRevC.96.024308,physics3040062,PhysRevC.104.L061306} treats the nucleus as an open quantum system through the inclusion of bound, resonant, and scattering states, which form the Berggren basis~\cite{BERGGREN1968265}. Furthermore, GSM has been generalized to the coupled-channel representation (GSM-CC)~\cite{PhysRevC.99.044606,Dong2017,PhysRevC.91.034609,PhysRevC.100.064303}, where reaction channels are constructed by coupling target with projectile states. GSM-CC allows the simultaneous calculations of nuclear structure and reaction cross sections within the same Hamiltonian. GSM-CC has been successfully applied to the $^{15}$O$(p,p)$~\cite{PhysRevC.106.L011301}, $^{18}$Ne$(p,p)$ ~\cite{PhysRevC.89.034624}, and $^{20}$Mg$(p,p)$~\cite{PhysRevC.111.034327} reactions.
%GSM-CC has been successfully applied to studies of low-energy elastic and inelastic scattering as well as radiative capture reactions, including the $^{15}$O$(p,p)$, $^{18}$Ne$(p,p)$, $^{6}$Li$(p,\gamma)^{7}$Be, and $^{6}$Li$(n,\gamma)^{7}$Li reactions.

%Experimentally, resonant elastic scattering reaction was performed to investigate the unbound $^{18}$Na~\cite{ASSIE2012198}. 
In this paper, GSM-CC is used for the investigation of the properties of $^{18}$Na. Its low-lying states, associated decay widths, as well as the cross section of the $^{17}\text{Ne}(p,p)$ reaction have been computed in the GSM-CC framework. Furthermore, the isospin asymmetry of the mirror pair $^{18}$Na/$^{18}$N has also been quantitatively assessed using the GSM-CC calculations.

\section{THE GAMOW SHELL MODEL IN THE COUPLED-CHANNEL REPRESENTATION}
The GSM is formulated in the Berggren basis~\cite{BERGGREN1968265}, which contains bound, resonance, and scattering states, and is generated by a finite depth potential. The one-body completeness is guaranteed for each Berggren partial wave of quantum numbers $l,j$: 
\begin{equation}\label{eq1}
\sum_{n}\left|u_{n}\right\rangle\left\langle u_{n}\right|+\int_{{L^{+}}}\left|u_{k}\right\rangle\left\langle u_{k}\right|dk=\mathbf{\hat{1}},
\end{equation}
where the $\left|u_{n}\right\rangle$ is a bound or resonance state, and $\left|u_k\right\rangle$ is a scattering state belonging to the complex contour $L^+$. All the discrete resonances present in the sum of Eq.(\ref{eq1}) must be encompassed in the $L^+$ contour.
The integration contour in Eq.(\ref{eq1}) is discretized by way of the Gauss-Legendre quadrature for computation purposes~\cite{Michel2009}. A many-body basis of Slater determinants can indeed be directly constructed from a discrete one-body basis.
% in analogy to the harmonic oscillator shell model method. The key difference is that the GSM Hamiltonian matrix is complex symmetric represented in the Berggren basis, and contains resonances and multibody scattering states bearing complex energies.
Nuclear states in GSM are determined by diagonalizing the Hamiltonian matrix with the Jacobi-Davidson and the overlap methods~\cite{Michel2009,PhysRevC.67.054311,MICHEL2020106978,Jacobi-davidson-methods}.

However, GSM is a structure tool and cannot be used to calculate reaction cross sections. Indeed, the many-body scattering states obtained by the diagonalization of the GSM matrix are generally complex linear combinations of many reaction channels. In particular, their asymptotic behavior is undefined.

To solve this problem, the target–projectile reaction channels are constructed in the GSM-CC~\cite{MichelSpringer,PhysRevC.99.044606,Dong2017,PhysRevC.91.034609,PhysRevC.89.034624}, where the target and projectile wave functions are generated by the GSM Hamiltonian. In this way, GSM-CC allows for the calculation of energy spectra and reaction cross sections using the same Hamiltonian.

The $A$-body resonant or scattering state of angular quantum numbers ($J_A, M_A$) is represented as linear combinations of reaction channels $c$ :
\begin{equation}\label{eq2}
\left\lvert \Phi_{M_{A}}^{J_A} \right\rangle = \sum_c \int_0^{+\infty} (\frac{u_c(r)}{r}) \left\lvert (c,r) \right\rangle r^2 dr.
\end{equation}
$u_c(r)$ is the radial amplitude of the channel $c$, where the radial coordinate $r$ represents the relative distance between the projectile and the core of the target. The channel state $\left\lvert (c,r) \right\rangle$ is defined by :
\begin{equation}\label{eq3}|(c, r)\rangle=\hat{\mathcal{A}}\left\{\left|\Psi_{\mathrm{T}}^{J_{\mathrm{T}}}\right\rangle \otimes|r \ell j\tau \rangle\right\}_{M_{\mathrm{A}}}^{J_{\mathrm{A}}},
\end{equation}
where the obtained $A$-body composite system, with total angular momentum $J_{\mathrm{A}}$ and projection $M_{\mathrm{A}}$, is generated by the antisymmetrized coupled product state of the $(A-1)$-body target state \(|\Psi^{J_T}_T\rangle\) and of a nucleon projectile state $|r \ell j\tau\rangle$, with $\tau$ representing the isospin quantum number of the nucleon (proton or neutron).

The GSM-CC equations are obtained by projecting the Schrödinger equation $\hat{H}|\Phi^{J_A}_{M_A}\rangle=E|\Phi^{J_A}_{M_A}\rangle$ onto each reaction channel $c$:
\begin{equation}\label{eq4}\sum_{c} \int_{0}^{+\infty}\left(H_{c c^{\prime}}\left(r, r^{\prime}\right)-E N_{c c^{\prime}}\left(r, r^{\prime}\right)\right) u_{c}\left(r^{\prime}\right) d r^{\prime}=0,
\end{equation}
with 
\begin{equation}\label{eq5}H_{c c^{\prime}}\left(r, r^{\prime}\right)=r r^{\prime}\left\langle(c, r)|\hat{H}|\left(c^{\prime}, r^{\prime}\right)\right\rangle,
\end{equation}
\begin{equation}\label{eq6}
N_{c c^{\prime}}\left(r, r^{\prime}\right)=r r^{\prime}\left\langle(c, r)|\left(c^{\prime}, r^{\prime}\right)\right\rangle.
\end{equation}

The matrix elements of the Hamiltonian and overlap kernels in Eqs.(\ref{eq5},\ref{eq6}) are calculated by taking heed of the completeness properties of the Berggren basis. The nucleon states \(|r \ell j \tau \rangle\) are thereby expanded with \(|n \ell j \tau \rangle\) Berggren basis states, so that the expression of the channel wave functions in Eq.(\ref{eq3}) becomes:
\begin{equation}\label{eq7}
\left\lvert (c,r) \right\rangle=\sum_{n} \frac{u_{n}(r)}{r}\left\{a_{n \ell j \tau}^{\dagger}\left|\Psi_{\mathrm{T}}^{J_{\mathrm{T}}}\right\rangle\right\}_{M_{\mathrm{A}}}^{J_{\mathrm{A}}}.
\end{equation}
As the $|\Psi^{J_T}_T\rangle$ target state and \(|n \ell j \tau \rangle\) projectile wave functions are represented in the same Berggren basis, the GSM-CC matrix elements are immediate to implement from shell model algebra~\cite{MichelSpringer}.

The antisymmetry requirement between target and projectile in the channels $\left\lvert (c,r) \right\rangle$ induces their non-orthogonality, which is conveyed by the overlap matrix of Eq.(\ref{eq6}), so that Eq.(\ref{eq4}) becomes a generalized eigenvalue problem, which reads in matrix form:
\begin{equation}\label{eq8}
\hat{H}U=E\hat{N}U,
\end{equation}
where $U=\{u_c(r)\}_c$ is the vector of channel radial functions. It is natural to reduce the former generalized eigenvalue problem to standard form by introducing the similar transformation $W=\hat{N}^{1/2}U$, and the modified Hamiltonian $\hat{H}_m=\hat{N}^{-1/2}\hat{H}\hat{N}^{-1/2}$:
\begin{equation}\label{eq9}
\hat{H}_mW=EW,
\end{equation}
where $W=\{w_c(r)\}_c$ is the reduced channel function to be determined.

By introducing the operator $\hat{\Delta}=\hat{\mathcal{N}}^{-\frac{1}{2}}-\hat{\mathbf{1}}$, the modified Hamiltonian $\hat{H}_m$ can be separated into long-range and short-range parts:
\begin{equation}\label{eq10}
H_{m}=(\Delta+\hat{\mathbf{1}}) H(\Delta+\hat{\mathbf{1}})=H+H \Delta+\Delta H+\Delta H \Delta .
\end{equation}
While its long-range part consists of the initial GSM Hamiltonian represented in the Berggren basis, the finite-range terms involving the $\Delta$ operator can be efficiently expanded with the harmonic oscillator basis.

The former modification allows the GSM-CC equations to be rewritten as an integro-differential coupled-channel system generated by the local $V^{(loc)}_c(r)$ and non-local $V^{(non-loc)}_{cc'}(r,r')$ coupled-channel potentials, which provide the reduced channel function $w_c(r)$ :
\begin{equation}\label{eq11}
\begin{aligned}& \left(\frac{\hbar^{2}}{2 \mu}\left(-\frac{d^{2}}{d r^{2}}+\frac{\ell_{c}\left(\ell_{c}+1\right)}{r^{2}}\right)+V_{c}^{(\mathrm{loc})}(r)\right) w_{c}(r) \\+ & \sum_{c^{\prime}} \int_{0}^{+\infty} V_{c c^{\prime}}^{(\mathrm{non}-\mathrm{loc})}\left(r, r^{\prime}\right) w_{c^{\prime}}\left(r^{\prime}\right) d r^{\prime} \\& =\left(E-E_{\mathrm{T}_{\mathrm{c}}}\right) w_{c}(r),\end{aligned}
\end{equation}
where ${\mu}$ is the reduced mass of the projectile, $l_c$ is the channel orbital momentum, $E_{\mathrm{T}_{\mathrm{c}}}$ is the target energy in the channel $c$. 

Eq.(\ref{eq11}) is highly non-local when channel coupling is strong. As a consequence, its direct integration in coordinate space by way of iterative methods such as the equivalent potential method~\cite{Michel2009,VAUTHERIN1967175} cannot be conveyed. To address this issue, a novel numerical technique has been developed that utilizes the completeness properties of the Berggren basis~\cite{MichelSpringer,PhysRevC.99.044606}. Eq.(\ref{eq11}) can indeed be transformed either into a matrix diagonalization problem for bound and resonance states, or to a linear system for scattering states.
The initial vector of the channel function $U=\{u_c(r)\}_c$ is recovered by the solution $W=\{w_c(r)\}_c$ of Eq.(\ref{eq11}) with the following expression:
\begin{equation}\label{eq12}
u_{c}(r)=w_{c}(r)+\sum_{c^{\prime}} \int_{0}^{+\infty} \Delta_{c c^{\prime}}\left(r, r^{\prime}\right) w_{c^{\prime}}\left(r^{\prime}\right) d r^{\prime}.
\end{equation}

\section{MODEL SPACE AND HAMILTONIAN}
In the GSM framework, the nucleus is considered as a picture of the core + valence nucleon, where $^{14}$O is used as the inner core in this work.
Since $^{17}$Ne and $^{18}$Na are proton-rich nuclei, it is sufficient for the neutron one-body space to consist in the harmonic oscillator shell $0p_{1/2}$, $0d_{5/2}$, $0d_{3/2}$ and $1s_{1/2}$.
Proton $spd$ partial waves are represented by the Berggren basis, which contains the $S$-matrix poles $0d_{5/2}$, $0d_{3/2}$ and $1s_{1/2}$ and complex contours of the $s_{1/2}$, $p_{1/2}$, $p_{3/2}$, $d_{3/2}$ and $d_{5/2}$ partial waves. The $f_{5/2}$ and $f_{7/2}$ partial waves are generated by the HO basis, because their larger orbital angular content, associated with a higher centrifugal barrier, results in weaker continuum effects. Not more than two protons are allowed to occupy scattering states.

To eliminate spurious center of mass (CM) excitations in the GSM wave functions, the Hamiltonian is expressed in the cluster orbital shell model (COSM) framework~\cite{PhysRevC.38.410}:
%In COSM coordinates, the nucleus is treated as a core plus valence particles, so the coordinates of valence particles are defined relatively to the center of mass of the core. The GSM Hamiltonian is: 
\begin{equation}\label{eq}
\hat{H}=\sum_{i=1}^{N_{val}}\left(\frac{\hat{\mathbf{p}}_i^2}{2\mu_i}+\hat{U}_{core}^{(i)}\right)+\sum_{i<j}^{N_{val}}\left(\hat{V}_{ij}+\frac{\hat{\mathbf{p}}_i\cdot\hat{\mathbf{p}}_j}{M_{core}}\right),
\end{equation}
where $N_{val}$ is the valence nucleons number, ${\mu_i}$ is the reduced mass of nucleon and ${M_{core}}$ is the core mass. The potential $\hat{U}_{core}^{(i)}$ represents the one-body mean field of the core acting on each valence nucleon, while $\hat{V}_{ij}$ is the two-body interaction between valence nucleons. The last term represents the recoil term.

%In the calculation of $^{17}$Ne, $^{18}$Na and $^{17}\text{Ne}(p, p)$, we use $^{14}$O as the inner core. 
%The Hamiltonian consists of the Woods-Saxon potential describing the $^{14}$O core and the effective two-body interaction among valence nucleons.
The interaction between the core and valence nucleons is mimicked by a Woods-Saxon (WS) potential with spin-orbit term. %The parameters are chosen to reproduce $^{17}Ne$ and $^{18}$Na spectra.
The WS core potential has a diffuseness $d= 0.65$ fm and a Coulomb radius $r_{\text{Coul}}=2.5$ fm. The WS radius is 2.98 fm for proton and 3.2 fm for neutron. The WS central potential depth $V_o$ is 51.5 MeV ($l=0$) or 49.75 MeV ($l>0$) for proton, and 50.5 MeV ($l=0,2$) or 49.5 MeV ($l=1$) for neutron. The WS spin-orbit potential depth $V_{so}$ is fixed at 6.5 MeV for all $l>0$ partial waves. 
The nucleon-nucleon interaction is described by the effective field theory (EFT) interaction~\cite{MACHLEIDT20111,RevModPhys.81.1773,RevModPhys.92.025004}. An $A$-dependence of the two-body interaction is imposed to account for the effect of three-body interactions, while the Coulomb interaction for protons is considered exactly. The parameters of EFT interaction are adopted from Refs.~\cite{PhysRevC.100.064303,hkmy-yfdk}.

In the GSM-CC calculation, the channel states of $^{18}$Na are constructed by coupling the ground state ($J^\pi = 1/2^-$) and the excited states ($J^\pi = 3/2^-$, $5/2^-$, $1/2^+$, $3/2^+$, $5/2^+$) of the target nucleus $^{17}$Ne with the proton projectile in all its partial waves bearing $l \leq 3$, where the target and projectile wave functions are provided by the GSM Hamiltonian.

\iffalse
\begin{table}[!htb]
\centering
\setlength{\tabcolsep}{1.5mm}
\begin{tabular}{ccccccc}
\toprule
Parameter & & & Proton & & & Neutron \\
\midrule
$R_0$ & & & 2.98 fm & & & 3.2 fm \\
$V_o(l=0)$ & & & 51.5 MeV & & & 50.5 MeV \\
$V_{so}(l=0)$ & & & 0 MeV & & & 0 MeV \\
$V_o(l=1)$ & & & 49.75 MeV & & & 49.5 MeV \\
$V_{so}(l=1)$ & & & 6.5 MeV & & & 6.5 MeV \\
$V_o(l=2)$ & & & 49.75 MeV & & & 50.5 MeV \\
$V_{so}(l=2)$ & & & 6.5 MeV & & & 6.5 MeV \\
$V_o(l=3)$ & & & 49.75 MeV & & & - \\
$V_{so}(l=3)$ & & & 6.5 MeV & & & - \\
\bottomrule
\end{tabular}
\caption{Parameters of the WS potential of the $^{14}$O core used in the GSM and GSM-CC calculation of $^{17}$Ne and $^{18}$Na.}
\label{WS potential}
\end{table}
\fi

\section{Results} 

%corrective factor
We firstly describe the spectrum of $^{17}$Ne befalling from a GSM calculation, thereby achieving good agreement between the calculated binding energy and experimental data. 
Subsequently, with all parameters unchanged, we conducted GSM-CC calculations for $^{18}$Na within that unified framework, whereby the GSM and GSM-CC Hamiltonians are in principle identical. However, different truncation schemes are imposed in these two approaches. To reconcile this discrepancy and optimize the GSM-CC results consistently with experimental data, the two-body matrix elements of the nuclear interaction are multiplied by corrective factors in the GSM-CC calculations. However, as they bear values close to unity, the introduced modification does not alter the fundamental physical properties of the Hamiltonian initially fitted within the GSM framework. In this work, the corrective factors are fixed to the values 0.989, 0.97, 0.992, and 1.078 in the channels of quantum numbers $J^\pi = 0^-, 1^-, 2^-, 3^-$, respectively. 

\begin{figure}[!htb]
\includegraphics[width=0.3\textwidth]{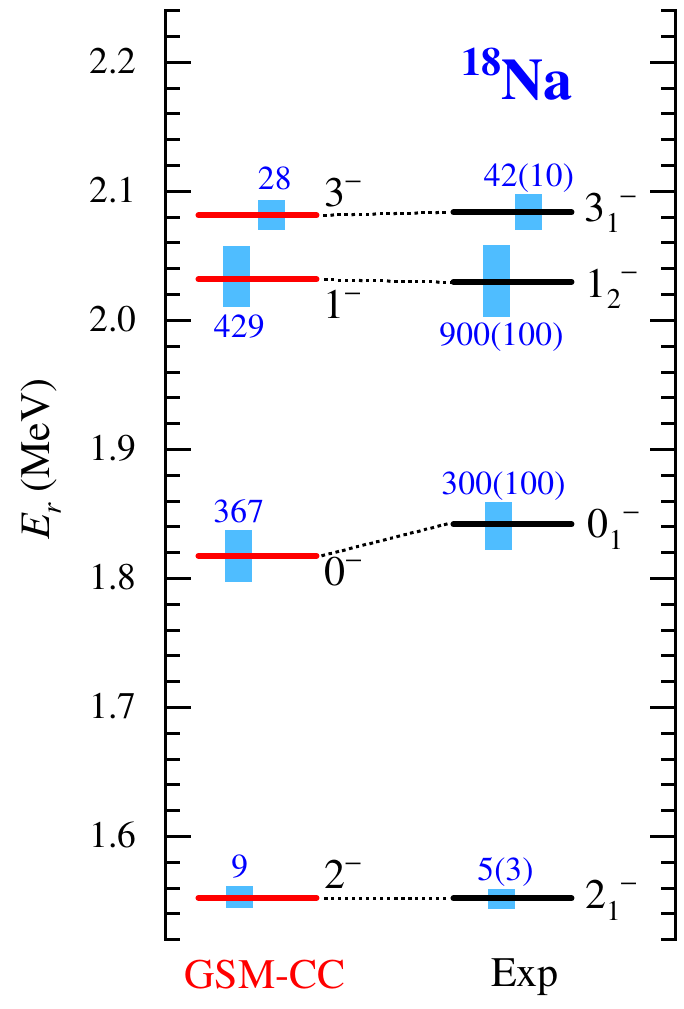}
\caption{\label{18Na}
Low-lying energy spectrum of the $^{18}$Na calculated by GSM-CC, along the experimental data (Exp).  Only states important for the description of cross sections are depicted. Excitation energies $E_r$ are provided in MeV relatively to the ground state of $^{17}$Ne. The calculated proton-emission width, in keV, is indicated by blue values next to the energy level. Experimental data are taken from Ref.~\cite{ASSIE2012198}.}
\end{figure}

%Fig.1
The obtained composite states and their corresponding energies in $^{18}$Na are shown in Fig.~\ref{18Na}. 
The excitation energy $E_r$ of $^{18}$Na is defined with respect to the ground state energy of $^{17}$Ne. 
The energies of the depicted low-lying states, which are determinant in cross section of interest, are very well reproduced in our calculations in Fig.~\ref{18Na}, with the largest discrepancy for the $0^-$ state which is only 35 keV lower than the experimental energy. 
Experimentally, the influence of the $1^-_1$, $2^-_2$ states of $^{18}$Na is very small~\cite{ASSIE2012198}, so that the latter states are not included in cross section evaluation. 
The decay widths are marked in blue next to the energy levels in the figure. The $2^-$ and $3^-$ states have very small widths, whereas the $1^-$ and $0^-$ states exhibit broader widths. 

Detailed decay branches and widths, along with the corresponding experimental data, are presented in Tab.~\ref{18Na-tab}. Theoretical values from previous SM calculations in Ref.~\cite{ASSIE2012198} are also provided for comparison. 
The $2^-$ state of $^{18}$Na is very narrow, with a calculated width of 9 keV, in excellent agreement with the experimental value of 5(3) keV and the SM predictions of 8 keV. The primary decay channel leads to the ground state $1/2^-$ of $^{17}$Ne, emitting a $d$-wave proton. 
The decay widths of the $0^-$ and $1^-$ states are relatively broad, around several hundred keV, with primary decays to the ground state of $^{17}$Ne via $s$-wave proton emission, and a small fraction decaying to the excited state $3/2^-$ via $d$-wave emission. The width of the $0^-$ state matches experimental data and SM predictions, while that of the $1^-$ state is slightly smaller in the GSM-CC calculation.
The $3^-$ state also has a narrower decay width, primarily decaying to the ground state of $^{17}$Ne with $d$-wave proton emission. 
Our calculated spectroscopic factors indicate that the $2^-$ and $3^-$ states predominantly exhibit couplings between the $^{17}$Ne$_{\rm{g.s.}}$ and the $d_{5/2}$ orbital, while the $0^-$ and $1^-$ states are primarily coupled with the $s_{1/2}$ orbital, as shown in Table~\ref{Spectroscopic factors}. The results obtained from both GSM and GSM-CC calculations are very close to the shell model (SM) calculations reported in Ref.~\cite{ASSIE2012198}.

\begin{table}[!htb]
\centering
\setlength{\tabcolsep}{0.26mm}
\begin{tabular}{cccccccccccc}
\hline\hline
\multirow{3}*{$J^\pi$}  &  \multicolumn{5}{c}{$\Gamma$ to $^{17}\text{Ne}_{\text{g.s.}}$(keV)} & \multicolumn{6}{c}{$\Gamma$ to $^{17}\text{Ne}_{3/2^-}^*$(keV)}\\
\cline{2-6} \cline{8-12}
 & Exp & Ref. & \multicolumn{3}{c}{GSM-CC} & & Exp & Ref. & \multicolumn{3}{c}{GSM-CC} \\
\cline{4-6} \cline{10-12}
 &  & ~\cite{ASSIE2012198} & $d_{5/2}$ & $d_{3/2}$ & $s_{1/2}$ &  &  & ~\cite{ASSIE2012198} & $d_{5/2}$ & $d_{3/2}$ & $s_{1/2}$ \\
\hline
$2^-$ & 5(3) & 8 & 9 & 0.006 & - & & <1 & 0 & - & - &- \\
$0^-$ & 300(100) & 189 & - & - & 367 & & <10 & 0 & - & <0.001 &- \\
$1^-$ & 900(100) & 1275 & - & 0.10 & 428 & & <100 & 4.8 & 0.72 & <0.001 & 0.06\\
$3^-$ & 42(10) & 31 & 28 & - & - & & <1 & 0.04 & <0.001 & <0.001 &- \\
\hline\hline
\end{tabular}
\caption{Partial decay widths of low-lying states in $^{18}$Na calculated with GSM-CC, compared with experimental data and theoretical calculations. Both the experimental results and the theoretical calculations are from Ref.~\cite{ASSIE2012198}.}
\label{18Na-tab}
\end{table}

\begin{table*}[!htb]
\centering
\setlength{\tabcolsep}{1.4mm}
\begin{tabular}{cccccccccccccccc}
\hline\hline
\multirow{2}*{$J^\pi$} & \multicolumn{4}{c}{$^{17}\text{Ne}_{\text{g.s.}}\otimes d_{5/2}$} & \multicolumn{4}{c}{$^{17}\text{Ne}_{\text{g.s.}}\otimes s_{1/2}$} &  \multicolumn{4}{c}{$^{17}\text{Ne}_{3/2^-}^*\otimes d_{5/2}$} & \multicolumn{3}{c}{$^{17}\text{Ne}_{3/2^-}^*\otimes s_{1/2}$} \\
\cline{2-4}\cline{6-8}\cline{10-12}\cline{14-16}
 & GSM & GSM-CC & SM & & GSM & GSM-CC & SM & & GSM & GSM-CC & SM & & GSM & GSM-CC & SM \\
\hline
$2^-$ & 0.701 & 0.737 & 0.644 & & - & - & - & & 0.140 & 0.201 & 0.311 & & 0.057 & 0.066  & 0.042 \\
$0^-$ & - & - & - & & 0.846 & 0.836 & 0.759 & & - & - & - & & - & -  & -\\
$1^-$ & - & - & - & & 0.842 & 0.835 & 0.654 & & 0.256 &  0.251 & 0.031 & & <0.001 & <0.001  & 0.027\\
$3^-$ & 0.707 & 0.733 & 0.621 & & - & - & - & & 0.073 & 0.075 & 0.109 & & - & -  & -\\
\hline\hline
\end{tabular}
\caption{Spectroscopic factors of low-lying states in $^{18}$Na calculated with GSM and GSM-CC. The shell-model spectroscopic factors from Ref.~\cite{ASSIE2012198} are also listed for comparison.}
\label{Spectroscopic factors}
\end{table*}

\begin{figure}[!htb]
\includegraphics[width=0.48\textwidth]{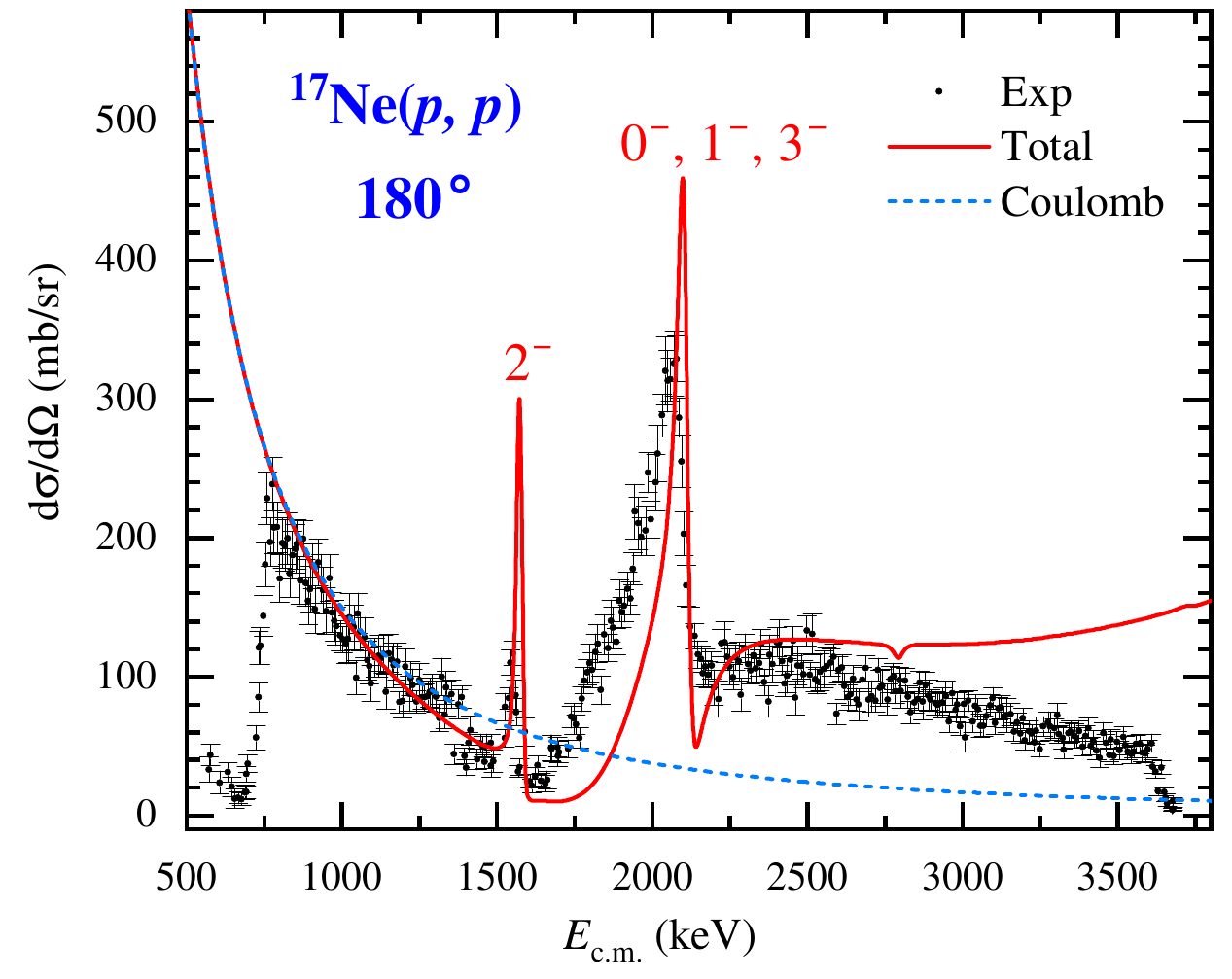}
\caption{\label{pp} {Excitation function of the $^{17}$Ne($p, p$) scattering reaction calculated with GSM-CC, along with the experimental data from Ref.~\cite{ASSIE2012198} for comparison. The solid red line represents the GSM-CC results using a Hamiltonian that includes both nuclear and Coulomb interactions. The dotted blue line corresponds to the calculation considering only the Coulomb interaction. Energies and cross sections are defined in the center of mass frame at 180 degrees.}}
\end{figure}

%Based on the previous analysis, the calculations were performed within the COSM framework, which involves approximations, while the experimental cross section is in the CM frame. Additionally, certain reaction channels may have been neglected in the GSM-CC Hamiltonian. These factors could lead to a decrease in energy when transitioning to the CM frame and cause a leftward shift in the cross section. To correct for this, we applied an energy shift factor of 0.981 for phenomenological adjustment. 
We evaluated the excitation function of the $^{17}\text{Ne}(p,p)$ scattering reaction, as shown in Fig.~\ref{pp}. 
While the Coulomb interaction predominates at low energies, several resonance peaks reflecting the structure of the composite nucleus $^{18}$Na occur at higher energies. The positions and widths of these peaks are related to the low-lying resonance states of $^{18}$Na (labeled next to their associated resonance peaks in Fig.~\ref{pp}).
In the projectile energy range of 800-2250 keV, our results are in good agreement with experimental data. 
The first peak corresponds to the narrow $2^-$ resonance, with a calculated energy of 1553 keV and a width of 9 keV. This is consistent with the experimental $2^-$ spin assignment at $E_r = 1.552(5)$ MeV and $\Gamma = 5(3)$ keV.
The second broad peak corresponds to the contributions of the $0^-, 1^-, 3^-$ states, as observed experimentally. It is somewhat narrower in GSM-CC than that of the R-matrix fit of Ref.~\cite{ASSIE2012198}, which is likely due to the small GSM-CC width of the $1^-$ state, resulting in a narrower peak.
Therefore, the GSM-CC method could successfully reproduce the resonance peaks observed in the experimental cross section. %, indicating that the model space used in this study accurately captures all key physical features and effectively reproduces the $^{17}\text{Ne}(p, p)$ reaction process
At high energies, discrepancies emerge between the calculated and experimental results. These deviations may be attributed to the missing reaction channels and core excitations.
The reaction channels involving many-particle projectiles (such as $^2$H, $^3$H, $^3$He, $^4$He) are indeed not included in our calculations. Added to that, excited states of the $^{14}$O core can be found 5-6 MeV above ground state energy, which are thus neglected in our model and could have an influence on results.

We will now report the results concerning isospin symmetry breaking in the mirror nuclei $^{18}$Na/$^{18}$N. They follow similar applications of GSM-CC in successful studies of the mirror pairs $^{16}$F/$^{16}$N in Ref.~\cite{PhysRevC.106.L011301}.
Since $^{18}$Na and $^{18}$N are mirror nuclei, the core, model space, and Hamiltonian for $^{18}$N are derived from those of $^{18}$Na by exchanging protons and neutrons. Specifically, the calculations involving $^{18}$Na demand $^{14}$O to be the inert core, whereas it is superseded by its mirror nucleus $^{14}$C in $^{18}$N. Correspondingly, the WS neutron potential parameters employed for $^{18}$N are adopted from the proton potential parameters of $^{18}$Na, and conversely for the WS core proton potential of $^{18}$N. This specific form of the Hamiltonian, incorporating proton-neutron exchange, enables the evaluation of the isospin symmetry breaking, as $^{18}$Na and $^{18}$N would share identical wavefunctions and energy spectra in the absence of Coulomb interaction.

\begin{figure}[!htb]
\includegraphics[width=0.49\textwidth]{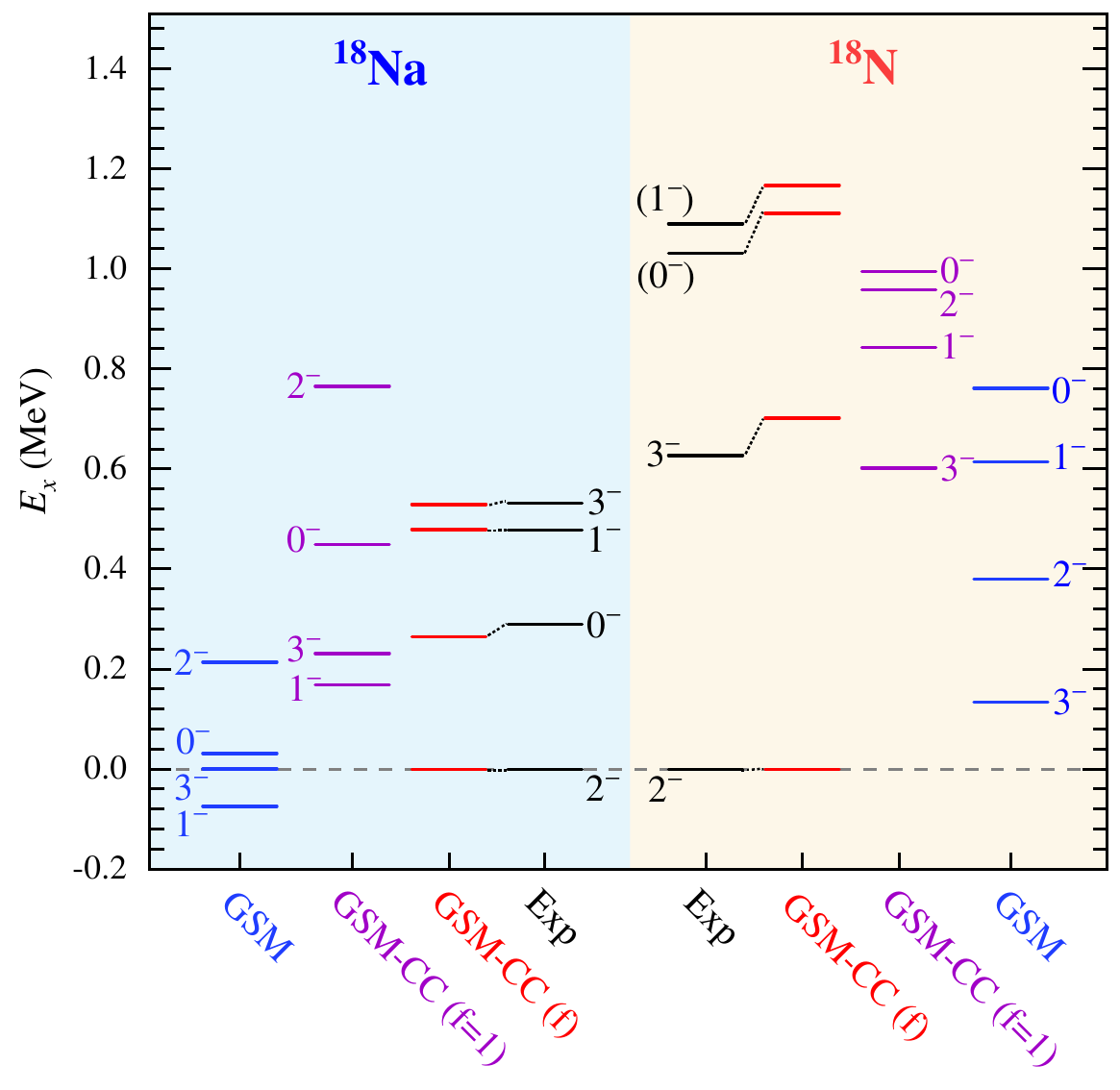}
\caption{\label{Na-N}  {Low-lying energy spectra of the $^{18}$Na and $^{18}$N nuclei obtained with GSM and GSM-CC (without and with corrections), along the experimental data (Exp). GSM-CC calculations without corrections are labeled by "f=1", while the "f" means the GSM-CC calculations with corrective factors. The energy spectra obtained from GSM and GSM-CC (f=1) calculations are referenced with respect to the 2$^-$ state energy calculated from GSM-CC (f). The 2$^-$ state energies of GSM-CC (f) and their experimental values are put to zero to be able to compare their mirror energy difference (MED). Excitation energies $E_x$ are given in MeV. Experimental data of $^{18}$N are taken from Ref.~\cite{PhysRevC.104.L041301}.}}
\end{figure}

In Fig.~\ref{Na-N}, we present the spectra of mirror nuclei $^{18}$Na/$^{18}$N obtained in GSM and GSM-CC calculations. In particular, both GSM-CC calculations with and without corrective factors are presented to highlight their impact on spectra. The GSM-CC results without corrective factors are labeled as GSM-CC (f=1), while the corrected calculations are labeled as GSM-CC (f). The spectra from GSM and GSM-CC (f=1) are shown relative to the energy of the $2^-$ state obtained from GSM-CC (f). The GSM calculations fail to reproduce the correct level ordering, while the $0^-$, $1^-$, and $3^-$ states are overbound. Even though GSM-CC (f=1) slightly ameliorates spectra overall, it still lacks the accuracy needed to satisfactorily reproduce both structure and reaction observables. The introduction of corrective factors in the GSM-CC (f) calculations is necessary to successfully reproduce both experimental level ordering and energies.

As previously discussed, we applied several corrective factors to match the experimental spectrum of $^{18}$Na. Interestingly, the same factors also successfully reproduce the spectrum of $^{18}$N containing the same states, as shown in the GSM-CC(f) results shown in Fig.~\ref{Na-N}.
The calculated separation energy of the first excited state, $S_n^*(2^-)$, is 2.705 MeV for $^{18}$N, while its experimental value is 2.713 MeV. Moreover, the calculated spectrum of $^{18}$N is approximately 250 keV less bound than experimental data. However, these slight differences do not significantly affect its physical properties, as the calculated states of $^{18}$N can be found below neutron separation energy, $S_n = 2.828(24)$ MeV~\cite{Wang2021}. 

The energies of the $2^-$ states in GSM-CC(f) and their experimental values are set to zero (reference states) in Fig.~\ref{Na-N} to analyze the isospin symmetry breaking of the mirror nuclei consistently.
The energy spectra of $^{18}$N and $^{18}$Na should be identical in the absence of the Coulomb interaction. 
However, the differences in neutron and proton numbers, along with the effects of Coulomb interaction and continuum coupling, can lead to a mirror energy difference (MED).
TES, the most striking consequence of isospin symmetry breaking, is clearly visible in Fig.~\ref{Na-N}, where the obvious different excitation energies and different level orderings co-exist in these two mirror partners, especially for the $0^-$ and $1^-$ states. Moreover, both experimental and calculated resonance widths of the $0^-$ and $1^-$ states in $^{18}$Na depict broad resonance widths (see Fig.~\ref{18Na}), while $^{18}$N remains well bound. 
%The states in $^{18}$N should be mirrors of those in $^{18}$Na, but the obvious different excitation energies and different level orderings exit in these two mirror partners (see Fig.~\ref{Na-N}), especially for the $0^-, 1^-$ states, which is pronounced Thomas-Ehrman shift(TES).

To further understand the origin of the TES in $^{18}$Na and $^{18}$N, the main configurations of the low-lying states are investigated with GSM-CC in Tab.~\ref{tab:4x13_table}. Strong single-particle configurations can be interpreted as the core of $^{17}$Ne/$^{17}$N in the ground state $1/2^-$ or excited states $3/2^-$ and  $5/2^-$ being coupled with a single proton or neutron particle. 
For $^{18}$Na, the primary configurations of the $2^-$ and $3^-$ states involve $^{17}$Ne coupled with a proton in the $d_{5/2}$ orbital, while for the $0^-$ and $1^-$ states, $^{17}$Ne is coupled with a proton in the $s_{1/2}$ orbital. 
Similarly, for $^{18}$N, the dominant configurations of the $2^-$ and $3^-$ states consist in the $^{17}$N ground state coupled with a single neutron in the $d_{5/2}$ orbital, whereas it is coupled to a single neutron in the $s_{1/2}$ orbital in the $0^-$ and $1^-$ states.
In fact, the components of $^{18}$Na and $^{18}$N are essentially identical, differing only in the mirror exchange of protons and neutrons.

\begin{table}[!htb]
\centering
\begin{tabular}{ccccccc}
\hline\hline
\multirow{2}*{$J^\pi$}  & \multicolumn{3}{c}{$^{18}$Na} & \multicolumn{3}{c}{$^{18}$N}\\
\cline{3-4} \cline{6-7}
 & & Coupled channel  & Prob. & & Coupled channel & Prob.\\
\hline
$2^-$ & & $^{17}\text{Ne}(1/2^-) \otimes d_{5/2}$ & 71\%  & & $^{17}\text{N}(1/2^-) \otimes d_{5/2}$ & 68\% \\
    & & $^{17}\text{Ne}(3/2^-) \otimes d_{5/2}$ & 16\% & & $^{17}\text{N}(3/2^-) \otimes d_{5/2}$ & 17\% \\
    & & $^{17}\text{Ne}(5/2^-) \otimes d_{5/2}$ & 7\% & & $^{17}\text{N}(5/2^-) \otimes d_{5/2}$ & 11\% \\
$0^-$ & & $^{17}\text{Ne}(1/2^-) \otimes s_{1/2}$ & 81\% & & $^{17}\text{N}(1/2^-) \otimes s_{1/2}$ & 72\% \\
    & & $^{17}\text{Ne}(5/2^-) \otimes d_{5/2}$ & 19\% & & $^{17}\text{N}(5/2^-) \otimes d_{5/2}$ & 27\% \\
$1^-$ & & $^{17}\text{Ne}(1/2^-) \otimes s_{1/2}$ & 74\% & & $^{17}\text{N}(1/2^-) \otimes s_{1/2}$ & 71\% \\
    & & $^{17}\text{Ne}(3/2^-) \otimes d_{5/2}$ & 16\% & & $^{17}\text{N}(3/2^-) \otimes d_{5/2}$ & 20\% \\
$3^-$ & & $^{17}\text{Ne}(1/2^-) \otimes d_{5/2}$ & 70\% & & $^{17}\text{N}(1/2^-) \otimes d_{5/2}$ & 68\% \\
    & & $^{17}\text{Ne}(5/2^-) \otimes d_{5/2}$ & 16\% & & $^{17}\text{N}(5/2^-) \otimes d_{5/2}$ & 21\% \\
\hline\hline
\end{tabular}
\caption{The main configurations in $^{18}$Na and $^{18}$N, calculated by the GSM-CC method, arise from the coupling of $^{17}$Ne and $^{17}$N with a single proton and neutron, respectively. The probabilities of these configurations for the low-lying states are presented.}
\label{tab:4x13_table}
\end{table}

%\sout{For the ground state $1/2^-$ of $^{17}$Ne, the main configurations are $\pi (0d_{5/2})^2 \otimes \nu (0p_{1/2})^1$, $\pi (1s_{1/2})^2 \otimes \nu (0p_{1/2})^1$, contributing 73\%, 15\% to the total wave function, respectively. While for $^{17}$N, the main configurations of the ground state are $\nu (0d_{5/2})^2 \otimes \pi (0p_{1/2})^1$, $\nu (1s_{1/2})^2 \otimes \pi (0p_{1/2})^1$, contributing 77\%, 12\%, respectively. The components of excited states ($3/2^-, 5/2^-$) in $^{17}$Ne and $^{17}$N are also very similar.}

\begin{figure}[!htb]
\includegraphics[width=0.4\textwidth]{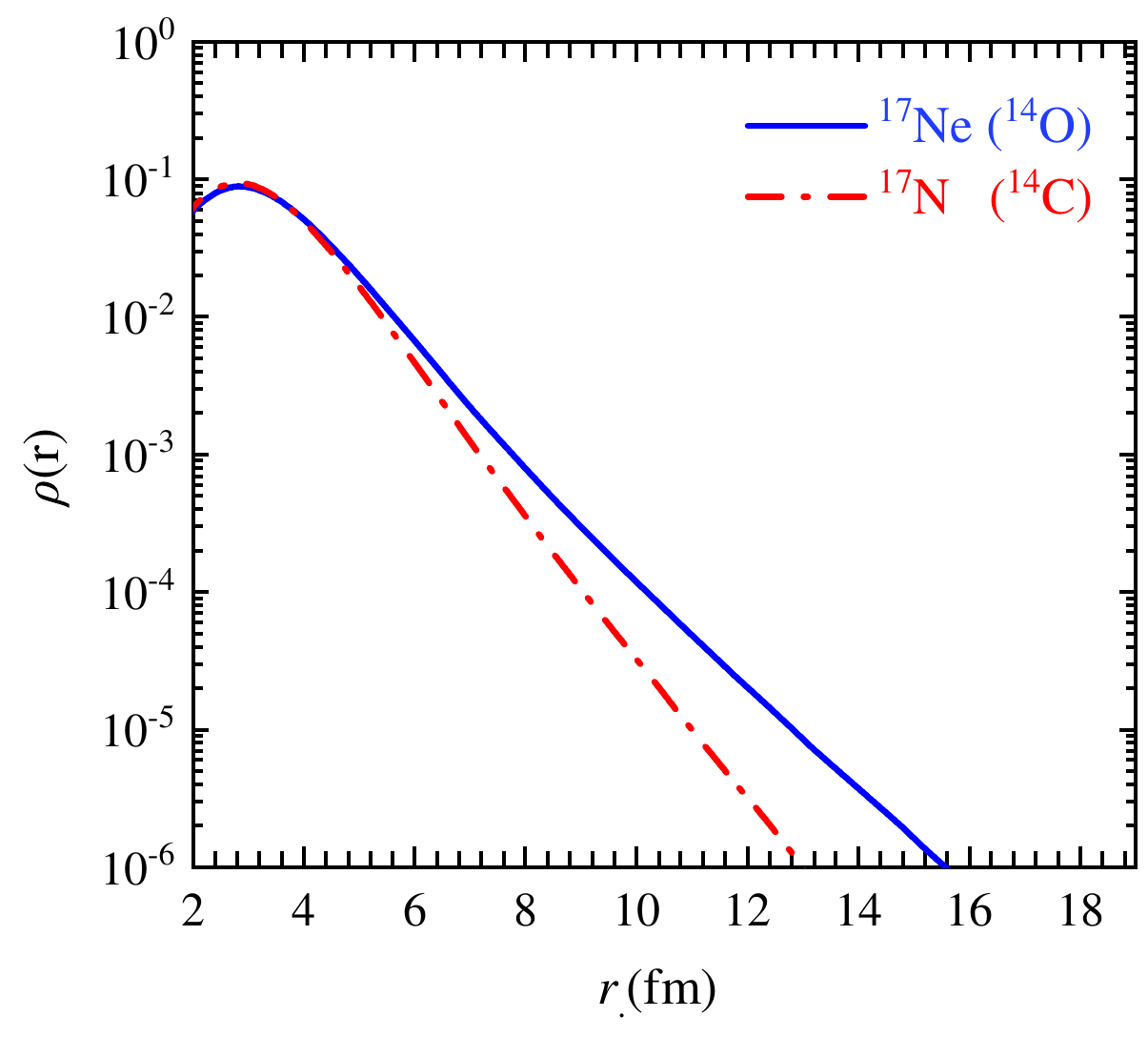}
\caption{\label{Ne-N}  {GSM density distributions of valence protons (neutrons) of the $^{17}$Ne ($^{17}$N) ground states, respectively. The used cores for both calculations are indicated in brackets. }}
\end{figure}

%Note that significant $s$-wave components occur in the $0^-$ and $1^-$ states, which differ from the primary configuration with $d$-wave in $2^-$ and $3^-$ states (see Tab.~\ref{tab:4x13_table}), for both $^{18}$Na and $^{18}$N.
Interestingly, in the mirror nuclei $^{16}$F/$^{16}$N, the $0^-$ and $1^-$ states are also dominated by the $s_{1/2}$ partial wave~\cite{PhysRevC.106.L011301,PhysRevC.108.064316,PhysRevC.90.014307}, exhibiting significant isospin symmetry breaking.
%Compared to those in $^{16}$N, the energies of the $0^-$ and $1^-$ states in $^{16}$F are significantly lower than the $2^-$ and $3^-$ states, and even the ground state of $^{16}$F is reversed. 
As there is no centrifugal barrier for $l = 0$, the wave function of the $s$-wave is more extended in space than in other partial waves. In the unbound nucleus $^{18}$Na, the presence of the Coulomb interaction and the more extended asymptotic behavior of the $s$ partial wave lead to a stronger coupling with the continuum states. This results in more binding energy and lower energies for the $0^-$ and $1^-$ states in $^{18}$Na, leading to significant isospin symmetry breaking between mirror nuclei $^{18}$Na/$^{18}$N. The same is true for the bound nucleus $^{16}$N and the unbound nucleus $^{16}$F. %In the mirror nuclei $^{16}$F and $^{16}$N, the Thomas-Ehrman shift is also attributed to the special role of the proton $s_{1/2}$ partial wave
We also calculated the density distributions of the mirror nuclei pair $^{17}$Ne/$^{17}$N, as shown in Fig.~\ref{Ne-N}.
Our results reveal that the valence protons density in $^{17}$Ne exhibits an extended spatial distribution compared to the valence neutrons density in $^{17}$N, supporting the two-proton halo properties in $^{17}$Ne. 
The nucleus $^{17}$Ne, located near the proton drip line, is then a promising candidate for exhibiting a two-proton halo structure due to its weakly bound valence protons and Borromean nature.
This observation also demonstrates the isospin symmetry breaking between the ground states of these mirror nuclei.

Based on the similar TES phenomenon in the mirror nuclei pairs $^{16}$F/$^{16}$N and $^{18}$Na/$^{18}$N, while Ref.~\cite{PhysRevC.106.L011301} also considered the GSM-CC method to study $^{16}$F/$^{16}$N, it is valuable to quantitatively discuss the isospin symmetry breaking in these mirror pairs. 
We can compare the continuum coupling effect through the different energy shifts of $0^-$ states reflecting the special role of its main component $s_{1/2}$ partial wave.
Following a previous idea, we fix the $2^-$ state as a reference state for energies. 
%Therefore, we can quantitatively discuss the effect of the coupling between the $s_{1/2}$ wave and the continuum state based on the magnitude of the energy shift of the $0^-$ state in the bound neutron-rich nuclei and the unbound proton-rich nuclei. 
According to the GSM-CC calculations in Ref.~\cite{PhysRevC.106.L011301}, compared to $^{16}$N, $E_x(0^-)$ in $^{16}$F is shifted downward by approximately 422 keV relative to the $2^-$ state, with corresponding experimental value around 530 keV.
The continuum coupling is stronger in the $0^-$ state than in the $2^-$ state, which leads to the $0^-$ state having a lower energy than the $2^-$ state in $^{16}$F, making the ground state reversal. 
In this work, $E_x(0^-)$ in $^{18}$Na is shifted downward by about 850 keV compared to $^{18}$N, with the experimental shift around 750 keV. The stronger continuum coupling in the $0^-$ state causes it to lie below the $3^-$ state in $^{18}$Na.
This level shift is also obtained in GSM and GSM-CC(f=1) calculations, albeit with a smaller value, being about 600 keV, compared to the GSM-CC(f) result and experimental data.
Similarly, recent experimental investigations have uncovered significant isospin symmetry breaking in the $^{20}$Al~\cite{hkmy-yfdk}, which is the isotones of $^{16}$F and $^{18}$Na. The combined experimental-theoretical analysis reveals the occurrence of ground-state inversion within the $^{20}$Al/$^{20}$N mirror nucleus pair.
Furthermore, it is clear that, compared to the mirror nuclei $^{16}$F/$^{16}$N, the extended $s$-wave in $^{18}$Na/$^{18}$N couples more strongly with the continuum, leading to a larger TES and more significant isospin symmetry breaking.
%Interestingly, in the mirror nuclei $^{16}$F and $^{16}$N, the Thomas-Ehrman shift is also attributed to the special role of the proton $s_{1/2}$ partial wave~\cite{PhysRevC.106.L011301,PhysRevC.108.064316,PhysRevC.90.014307}, where the energies of the $0^-$ and $1^-$ states in $^{16}$F are significantly lower than the $2^-$ and $3^-$ states, compared to those in $^{16}$N. 
%These states contain significant s-wave components, and the extended s-wave wavefunction enhances coupling with the continuum states, leading to additional binding energy. This results in a notable decrease in the $0^-$ and $1^-$ state energies in $^{16}$F compared to the $2^-$ and $3^-$ states.

% consistent with the general trend observed for nuclei in the sd−shell exhibiting large MED

\section{Summary}
Nuclei beyond the proton drip line exhibit exotic properties and have attracted significant interest due to their diverse proton-emission modes. The unbound nucleus $^{18}$Na, serving as an intermediate state in the sequential decay of $^{19}$Mg, has been explored experimentally, yet requires further in-depth theoretical investigation.
In this work, the GSM-CC, which incorporates continuum coupling effects, is applied to the $^{17}$Ne$(p, p)$ reaction. Additionally, isospin symmetry breaking is examined in the mirror pair formed by $^{18}$Na and $^{18}$N.

The obtained low-lying spectra of in $^{18}$Na and the excitation function of the $^{17}\text{Ne}(p,p)$ reaction are reproduced by the GSM-CC calculations. The first narrow peak in the excitation function corresponds to the $2^-$ resonance, and the second peak arises from the combined contributions of the $1^-$, $0^-$, and $3^-$ states, in good agreement with experimental observations. Detailed decay widths and branches indicates that the $2^-$ and $3^-$ states are very narrow, primarily decaying to the ground state of $^{17}$Ne via $d$-wave proton emission, while the broader $1^-$ and $0^-$ states decay mainly through $s$-wave emission. 

The isospin symmetry breaking, induced by both the Coulomb interaction and coupling to the continuum, is evident in the GSM-CC spectra of $^{18}$Na and $^{18}$N. The Thomas-Ehrman shift is clearly visible, particularly for the $0^-$ and $1^-$ states, due to the special role of their dominant $s_{1/2}$ partial wave. This is similar to the TES phenomenon observed previously in $^{16}$F and $^{16}$N.
Furthermore, compared to the mirror nuclei $^{16}$F/$^{16}$N, the extended $s$-wave in $^{18}$Na/$^{18}$N couples more strongly with the continuum, resulting in a larger TES and more significant isospin symmetry breaking. 

%Overall, the GSM-CC method successfully reproduced the experimental data, both in nuclear structure of $^{18}$Na and the $^{17}$Ne$(p, p)$ reaction cross sections, as well as the isospin asymmetry in the mirror pair $^{18}$Na and $^{18}$N.

\section{Acknowledgments}
This work has been supported by the National Key R\&D Program of China under Grant Nos. 2024YFE0109800, 2024YFE0109802, and 2023YFA1606403;
The National Natural Science Foundation of China under Grant Nos.  12205340, 12175281, 12347106, 12405141, 12475128, 12322507, and 12121005. 
The numerical calculations in this paper have been done on Hefei advanced computing center.

\bibliography{Ref}      
%\end{CJK}
\end{document}